# MoB$_2$: a new multifunctional transition metal diboride monolayer

Yipeng An[1,2,5], Shijing Gong[3,4], Yusheng Hou[2], Jie Li[2], Ruqian Wu[2,5], Zhaoyong Jiao[1], Tianxing Wang[1] and Jutao Jiao[1]

[1]School of Physics & International United Henan Key Laboratory of Boron Chemistry and Advanced Energy Materials, Henan Normal University, Xinxiang 453007, China

[2]Department of Physics and Astronomy, University of California, Irvine, California 92697, USA

[3]Key Laboratory of Polar Materials and Devices, Ministry of Education & Department of Optoelectronics, East China Normal University, Shanghai 200062, China

[4]Collaborative Innovation Center of Extreme Optics, Shanxi University, Taiyuan, Shanxi 030006, China

[5]Author to whom any correspondence should be addressed.

E-mail: ypan@htu.edu.cn (Y An) and wur@uci.edu (R Wu)

**Abstract**

Several layered transition metal borides can now be realized by a simple and general fabrication method [Fokwa et al., *Adv. Mater.* **2018**, *30*, 1704181], inspiring our interest to transition metal borides monolayer. Here, we predict a new two-dimensional (2D) transition metal diboride MoB$_2$ monolayer (ML) and study its intrinsic mechanical, thermal, electronic, and transport properties. The MoB$_2$ ML has isotropic mechanic properties along the zigzag and armchair directions with a large Young's stiffness, and has an ultralow room-temperature thermal conductivity. The Mo atoms dominate the metallic nature of MoB$_2$ ML. It shows an obvious electrical anisotropy and a current-limiting behavior. Our findings suggest that MoB$_2$ ML is a promising multifunctional material used in ultrathin high-strength mechanical materials, heat insulating materials, electrical-anisotropy-based materials, and current limiters. It is helpful for the experimentalists to further prepare and utilize the transition metal diboride 2D materials.

# 1. Introduction

Recently, atomically thin two-dimensional (2D) layered materials have aroused greast research interest because of their great potential for technological applications. Many 2D materials have been theoretically predicted and a number of them has been experimentally prepared, including graphene [1], hexagonal BN [2], silicene [3], transition metal dichalcogenides [4], phosphorene [5], germanene [6], MXene [7], borophene [8], and others [9-11]. Most of 2D materials show unique electronic, transport and optoelectronic properties and new branches of science emerge along with intensive interdisciplinary research activities [12-16]. It is foreseeable that further development of these materials will allow new technological innovations in a variety of areas such as lithium-sulfur batteries [17], negative differential resistive devices [18, 19], photodetectors [20], spin-filters [21], as well as spin field effect transistors [22], to name a few.

Transition metal boride layered materials as a particular class of materials are still underexplored even though some of their excellent properties such as superconductivity ($MgB_2$), mechanical toughness ($TiB_2$), and unusual conductivity ($ZrB_2$) have been recognized [23, 24]. Recent investigations have focused getting their monolayer structures. For instance, several monolayers of transition metal borides (such as $MgB_2$, $ZrB_2$, $TiB_2$, $CrB_4$, $MnB_6$, and $TiB_4$) [25-30] have been theoretically predicted, and some of them have been successfully fabricated in experiments [31]. More recently, simple and general synthetic routes were developed for fabricating various layered molybdenum borides [32]. Obviously, systematic fundamental studies are essential for revealing more exotic properties of molybdenum borides and related materials in the ultrathin regime. Especial for the $MoB_2$, which is composed of alternate planar Mo and B sheets, it is expected to prepare its monolayer structure by means of exfoliation method such as micromechanical cleavage[1], mechanical force-assisted liquid exfoliation [33], and ion intercalation-assisted liquid exfoliation [34].

In this report, we predict a new transition metal diboride $MoB_2$ ML by using a first-principles method, and confirm its dynamic and thermal stabilities. We find that the $MoB_2$ ML has isotropic mechanic properties along the zigzag and armchair directions with a large Young's stiffness. Furthermore, we uncover its intrinsic thermal transport, electronic, and electronic transport properties. The $MoB_2$ ML has ultralow thermal conductivity and exhibits a metallic nature due to the dominating contribution from Mo atoms. In addition, it shows the obvious electrical anisotropy and interesting current-limiting effects.

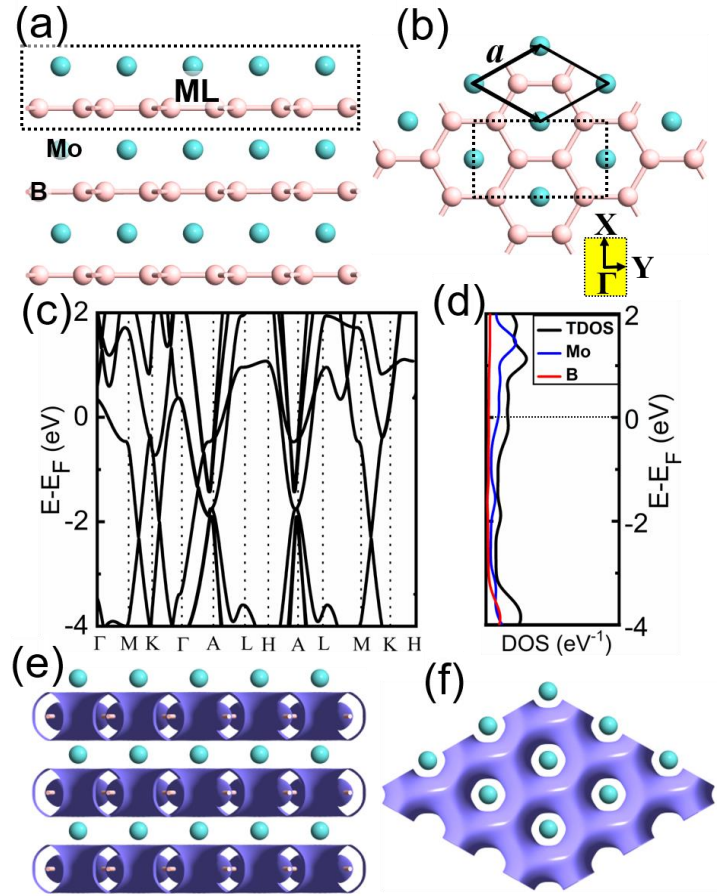

**Figure 1.** (a) Side and (b) top views of layered $MoB_2$ bulk material. The dotted box in (a) refers to the $MoB_2$ monolayer. (c) Band structures and (d) PDOS of $MoB_2$ bulk. (e) Side and (f) top views of 3D ELF of $MoB_2$ bulk at an isovalue of 0.45.

## 2. Methods

The atomic positions were optimized through density functional theory (DFT) as implemented in the Vienna *ab initi*o simulation package (VASP) [35]. The electron exchange-correlation effect was treated using the GGA-PBE [36]. A kinetic energy cut-off 500 eV was adopted with a 21 × 21 × 1 Γ-centered Monkhorst–Pack *k*-point grids, and a 20 Å vacuum along z-direction was used to eliminate interaction with other neighboring layers. The structure relaxation of MoB$_2$ were done until the total energy tolerance is less than $10^{-8}$ eV and the residual force on each atom is below $10^{-6}$ eV Å$^{-1}$, respectively.

The phonon spectra were calculated by the density functional perturbation theory approach, which are calculated based on the VASP through the PHONOPY [37] code. The thermal stability of MoB$_2$ ML was examined through *ab initio* molecular dynamics (AIMD) simulations based on the Nosé-Hoover method, and the simulation time lasts for 5 ps with a step of 1fs. The thermal transport properties of MoB$_2$ ML were obtained by solving the phonon Boltzmann transport equation and it is implemented in the ShengBTE code [38]. The acquired harmonic and anharmonic interatomic force constants(IFC) were generated by PHONOPY and THIRDORDER.py codes, respectively. The dynamic, thermal stability, and thermal transport properties of MoB$_2$ ML were confirmed with a 5 × 5 × 1 supercell containing 75 atoms in total.

The electronic transport properties of MoB$_2$ ML were calculated by combinig the DFT with the nonequilibrium Green's function approach (DFT-NEGF), as implemented in the Atomistix Toolkit package [39-41]. The core electrons of Mo and B were represented by the optimized Norm-Conserving Vanderbilt pseudo-potentials, and their valence states were described by linear combinations of atomic orbitals (LCAO) with SG15 [42] pseudo-potentials and basis sets. We used a 1 × 9 × 100 Monkhorst-Pack *k*-point grid-mesh to sample the 2D Brillouin zone of each electrode in the electronic transport calculations. Note that the spin-polarized calculations were performed to examine the ground-state configuration of $MoB_2$ ML, which is a nonmagnetic metal with weak spin-orbital coupling (SOC) effect.

## 3. Results and discussion

Figures 1(a) and (b) show the side and top views of layered $MoB_2$ bulk structure, which has the *P6/mmm* space group (No. 191) and consists of alternate planar Mo and B sheets. Our calculated lattice constant of its hexagonal primitive cell (containing one Mo and two B atoms) is $a = b = 3.03$ Å, and $c = 3.34$ Å, respectively, in consistent with the experimental data [43]. The lengths of B-B, Mo-Mo, and Mo-B bonds are 1.75, 3.03, and 2.42 Å, respectively. $MoB_2$ shows metallic characteristic (see figure 1(c)) in its crystalline phases. This feature mostly results from the Mo orbitals, according to the projected density of states (PDOS) as shown in figure 1(d). Figures 1(e) and (f) show the 3D electron localization function (ELF) of $MoB_2$ ML. High electron density clearly locates around the B-B bonds and the B atoms, indicating the strong covalent bonds that hold a 2D network together. On the contrary, very rare electronic states is founded among Mo atoms, and there is no strong covalent bonds formed between B and Mo atoms either. So, the B-B bonds are the major frameworks of $MoB_2$ bulk to resist the external compression, and it is easy to separate the layered $MoB_2$ such as by means of mechanical exfoliation, which is common used to prepare the graphene sheet [1].

Next, we mainly focus on the properties of 2D $MoB_2$ ML, which has the *P6mm* space group (No. 183), as illustrated in figures 1(a) and (b). We systematically investigate its structural stability, electronic structure, and electronic transport properties. After geometric optimization, the lattice constant *a* of hexagonal $MoB_2$ ML (see figures 1(a) and (b)) shrinks to 2.91 Å, and the lengths of B-B, Mo-Mo, and Mo-B bonds become 1.68, 2.91, and 2.31 Å, respectively. Its structure stability can then be ascertained from the following three aspects, i.e., cohesive energy, phonon spectra, and *ab initio* molecular dynamic (AIMD) simulations.

**Cohesive Energy.** The cohesive energy is obtained by the equation $E_{coh} = (E_{Mo} + 2 \times E_B - E_{MoB2})/3$ [44], where $E_{Mo}$, $E_B$, and $E_{MoB2}$ refer to the free-energies of Mo, B atom, and the 2D $MoB_2$ unit cell, respectively. The cohesive energy of $MoB_2$ ML is 7.21 eV per atom. This value is close to that of graphene (7.8 eV per atom), and is obviously greater than those of silicene (3.98 eV per atom) [44] and germane (3.26 eV per atom) [45], respectively, demonstrating the energetical stability of 2D $MoB_2$ ML.

**Dynamic Stability.** The dynamic stability of 2D $MoB_2$ ML is then examined with phonon calculations. A $5 \times 5 \times 1$ supercell (containing 75 atoms) is used to remove the constraints of periodic boundary condition. The phonon spectra have no imaginary phonon modes (see figure 2(a)), implying that the 2D $MoB_2$ ML is locally stable without any dynamic instability. Because of the existence of three atoms in each primitive cell, there are nine phonon branches, including three acoustic (A) branches and six optical (O) branches. The three acoustic branches stem from the vibrations of Mo atoms, while the optical branches from the contributions of B atoms, according to the projected phonon density of states (see figure 2(a)). These can be understood easily from a diatomic linear chain model, where acoustic (optical) branches mostly stem from the vibrations of the larger (smaller) mass. The highest vibrational frequency of acoustic branches is 6.1 THz, smaller than that of transition metal diboride (i.e., $MoS_2$) [46], and the highest (lowest) vibrational frequency of optical branches is 28.8 (10.1) THz, respectively.

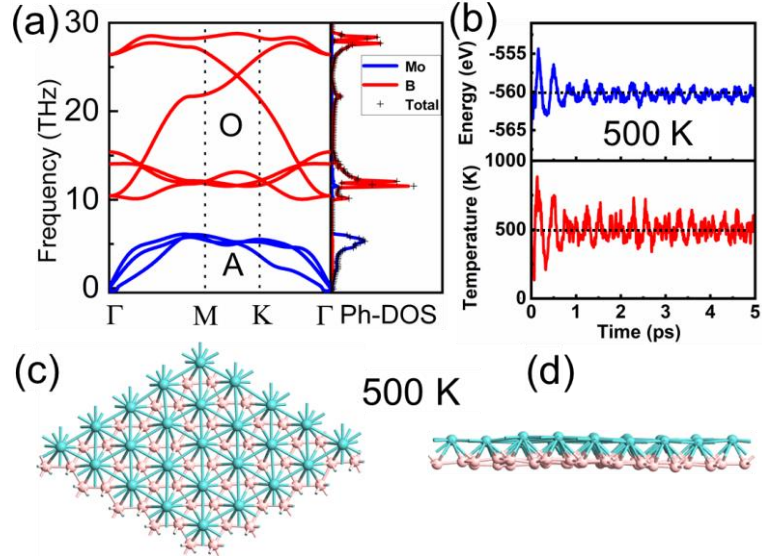

**Figure 2.** (a) Phonon spectra and projected phonon density of states. (b) Evolution of free energy and temperature *versus* simulation time in AIMD simulations at 500K. (c) Top and (d) side snapshots of a (5 × 5 × 1) supercell of MoB$_2$ ML after a 5 ps AIMD simulation.

**Thermal Stability.** Additionally, another key factor to check structure stability is its thermal stability at an elevated temperature. To this end, AIMD simulations are performed with a 5 × 5 × 1 supercell of MoB$_2$ ML. The AIMD simulations under a constant temperature (i.e., 500 K) and volume (NVT) ensemble are carried out for 5 ps with 1 fs time step. The evolution of the total free energy and temperature *versus* time is shown in figure 2(b), which shows that the total free energy has a slight oscillation at a constant value during the entire simulation. Moreover, the structure of the MoB$_2$ ML at the end of the simulation is given in figures 2(c) and (d), which keeps quite original planarity without obvious structural distortion at 500 K and starts to collapse at 750 K, strongly suggesting its good thermal stability.

**Mechanical Properties.** We further examine the mechanical stability of MoB$_2$ ML by calculating its elastic constants (ECs). The derived ECs of $c_{11}$, $c_{12}$, $c_{22}$, and $c_{66}$ are 225.8, 51.3, 225.8, and 87.2 N m$^{-1}$, respectively. Due to the hexagonal symmetry, it has $c_{11} = c_{22}$ and the additional relation that $c_{66} = (c_{11} - c_{22})/2$. The layer modulus $\gamma = (c_{11} + c_{22} + 2c_{12})/4$ is 138.5 N

m$^{-1}$, smaller than that of graphene (206.7 N m$^{-1}$) and hexagonal-BN (*h*-BN) [47]. The isotropic 2D Young's modulus (in-plane stiffness) for strains $Y = (c_{11}c_{22} - c_{12}^2)/c_{11}$ is 214.1 N m$^{-1}$, much larger than that of phosphorenes [48], while smaller than that of graphene and *h*-BN. Taking the value of 3.34 Å (i.e., the lattice length of MoB$_2$ bulk along *c* direction) as the thickness of MoB$_2$ ML, the calculated Young's stiffness value is 641 GPa, which is close to that of graphene (1029 GPa) [49] but is much larger than that of phosphorenes (164 GPa). Moreover, the Poisson's ration $\upsilon = c_{12}/c_{11}$ is 0.23, larger than that of graphene and close to that of *h*-BN. The 2D shear modulus $G$ (= $c_{66}$) is 87.2 N m$^{-1}$, smaller than that of graphene and *h*-BN. All these demonstrate that the MoB$_2$ monolayer has good mechanical properties and could have potential applications in ultrathin high-strength mechanical materials.

**Thermal Transport.** We now turn our attention to the thermal transport properties of MoB$_2$ ML. From the harmonic and anharmonic IFCs, the intrinsic lattice thermal conductivity ($\kappa_L$) of MoB$_2$ ML can be obtained by solving the linearized phonon Boltzmann equation within the single-mode relaxation time approximation (RTA) or iterative self-consistent (ITS) methods. Figure 3(a) displays the $\kappa_L$ obtained with both RTA and ITS methods from 100 to 700 K. We can see that both methods give consistent results in the whole temperature range. We choose the results from the ITS method in the following discussion. The $\kappa_L$ decreases as the temperature increases and can be well described by the T$^{-1}$ curve. The $\kappa_L$ of MoB$_2$ ML at room temperature is 11.54 W m$^{-1}$ K$^{-1}$ with a thickness of 20 Å, and the corresponding thermal sheet conductance is 23.08 nW K$^{-1}$, which is two (three) orders of magnitude lower than that of MoS$_2$ (graphene) [50, 51]. The ultralow $\kappa_L$ suggests that the MoB$_2$ ML could be a promising heat insulating material.

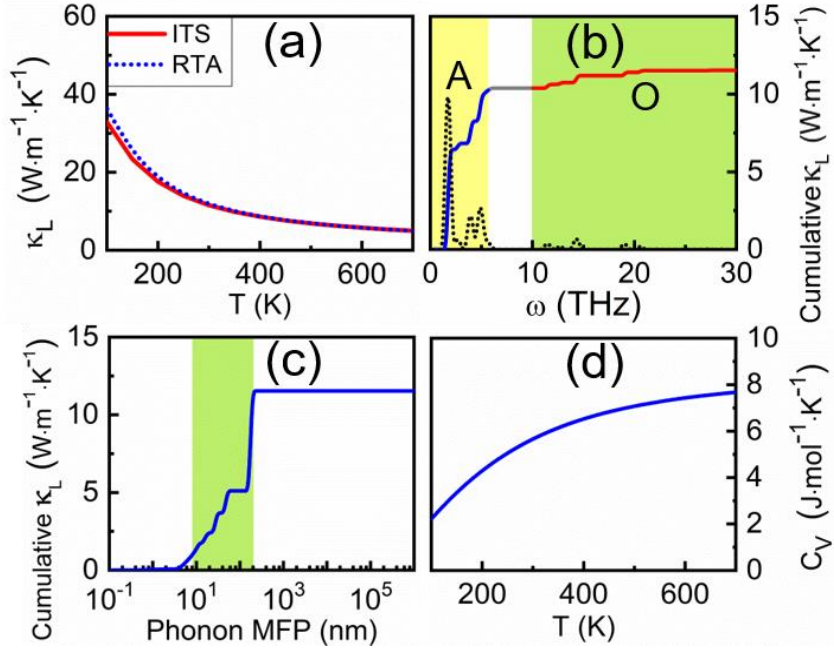

**Figure 3.** (a) Intrinsic $\kappa_L$ *versus* temperature. (b) Frequency-resolved cumulative $\kappa_L$ and the derivatives (dotted line) at 300 K. The yellow and green regions indicate the contribution from the acoustic(A) and optical(O) phonons, respectively. (c) Cumulative $\kappa_L$ with respect to phonon mean free path at 300 K. (d) Temperature-resolved specific heat capacity at constant volume.

To evaluate the relative contributions of the acoustic and optical branches to the total $\kappa_L$, the frequency-resolved cumulative $\kappa_L$ and its derivatives at room temperature (300 K) are depicted in figure 3(b). The acoustic and optical branches of MoB$_2$ ML provide a contribution of 90 and 10%, respectively. This is in consist with the common picture that high-frequency optical phonons have very little contribution to the $\kappa_L$. The derivatives suggest that it has a maximum contribution to the $\kappa_L$ at 1.73 THz. The size effect can be examined from the cumulative $\kappa_L$ with respect to phonon mean free path (MFP), as shown in figure 3(c). The MFP range of phonons that contribute significantly to $\kappa_L$ is about 8 to 200 nm. Phonons with the MFP larger than 200 nm give very little contribution to the $\kappa_L$, and the cumulative $\kappa_L$ approaches the maximum value 11.54 W m$^{-1}$ K$^{-1}$, which is equal to the room temperature thermal conductivity. Figure 3(d) shows the temperature-resolved specific heat capacity at constant volume ($C_V$). As the temperature increases, the $C_V$ reaches a saturation value 8.0 J mol$^{-1}$ K$^{-1}$, smaller than that of graphene (21 J mol$^{-1}$ K$^{-1}$) [52]. The $C_V$ at room temperature is 5.7 J mol$^{-1}$ K$^{-1}$, close to that of graphene (7 J mol$^{-1}$ K$^{-1}$).

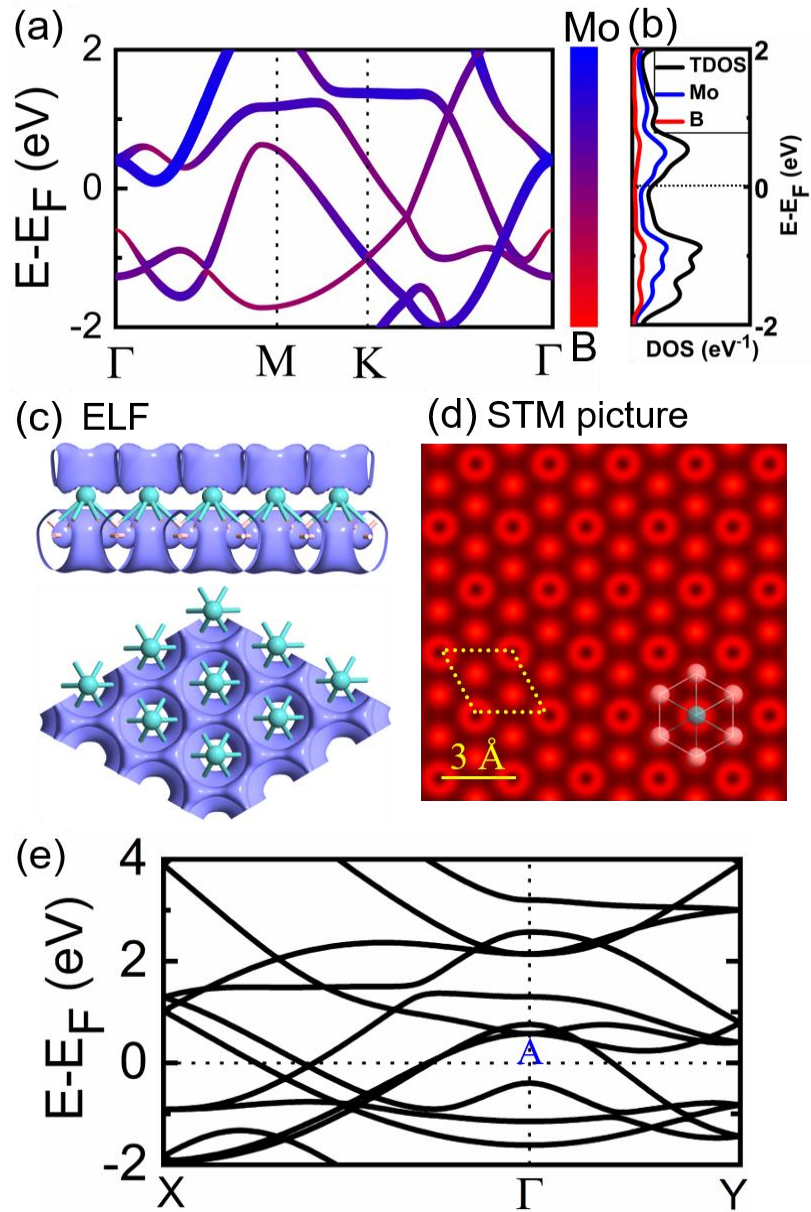

**Figure 4.** (a) Projected band structures and (b) PDOS of MoB$_2$ ML. (c) Side and top views of the 3D ELF of MoB$_2$ ML at an isovalue of 0.45. (d) Simulated STM picture of MoB$_2$ ML. (e) Bands of the rectangular unit cell along the $X$–$\Gamma$–$Y$ path. The Fermi level is shifted to zero.

**Electronic Structures.** Figures 4(a) and (b) show the projected band structures and density of states (PDOS) of hexagonal $MoB_2$ ML. It can be seen that the $MoB_2$ ML remains a metallic character in its bulk phase. This is mainly contributed by states of Mo atoms, according to the PDOS shown in figure 4(b). Figure 4(c) plots the side and top views of 3D ELF of $MoB_2$ ML, respectively. As a $MoB_2$ ML is separated from its layered bulk phase, additional strong metallic bonding appears between the Mo atoms. To supply more structural information for experimentalists, we simulate the scanning tunneling microscopy (STM) picture of $MoB_2$ ML, as shown in Figure 4(d). The bright hexagonal positions indicate the B atoms, and the others with a black hole in the honeycomb refer to the Mo atoms.

2D honeycomb monolayers often have distinguishing electronic and transport properties along the zigzag and armchair paths. For example, graphene presents a metallic feature along the zigzag direction, but shows a semiconducting character along the armchair one [53]. To examine the electrical anisotropy of $MoB_2$ ML, we depict its bands of rectangular unit cell along the $X$–$\Gamma$–$Y$ path in figure 1(b). The zigzag (i.e., along $\Gamma$–$X$) and armchair directions (i.e., along $\Gamma$–$Y$) are labeled as z-$MoB_2$ and a-$MoB_2$, respectively. Although $MoB_2$ ML shows metallic feature along the both directions, the difference is significant because more bands cross the Fermi level ($E_F$) along the $\Gamma$–$X$ than along the $\Gamma$–$Y$. This suggests that the z-$MoB_2$ could have better electroconductivity than a-$MoB_2$. Note that the $MoB_2$ ML is nonmagnetic according to the spin-polarized calculations. The spin-orbital coupling effect is insignificant for the $MoB_2$ ML, because it only results in tiny band splits and would not significantly change its intrinsic metallic properties.

**Electronic Transport Properties**. To further unveil the electrical properties of $MoB_2$ ML, we design a two-probe structure of z-$MoB_2$ and a-$MoB_2$ (see figures 5(a) and (b)), and directly calculate their electronic transport properties, including the transmission spectra (see figures 5(c) and (d)) and the current-voltage (*I*−*V*) curves (see figure 5(e)). This two-probe structure has a periodicity perpendicular to the transport direction between the left electrode (L) and the right electrode (R). The third direction is a slab with a vacuum of 20 Å. Each of the L and R electrodes is represented by a supercell, whose length along the transport direction is semi-infinite. The central scattering region (C) consists of 15 (10) unit cells for z-MoB$_2$ (a-MoB$_2$), which is five-times as long as electrode length and gives little influence to the transport. As a bias $V_b$ is applied across the electrodes, their Fermi energies are shifted accordingly. A forward bias generates an electric current from the L electrode to the R electrode, and vice versa. In this work, the current *I* through the z-MoB$_2$ and a-MoB$_2$ diode structures is calculated from the Landauer–Büttiker formula [54]

$$I(V_b) = \frac{2e}{h}\int_{-\infty}^{\infty}T(E,V_b)[f_L(E-\mu_L)-f_R(E-\mu_R)]dE, \qquad (1)$$

Here, $T(E,V_b)$ refer to the bias-resolved transmission coefficient, determined by the Green's functions; $f_{L/R}$ indicate the Fermi-Dirac distribution functions of the L/R electrodes; $\mu_L$ (= $E_F - eV_b/2$) and $\mu_R$ (= $E_F + eV_b/2$) are the electrochemical potentials of the L and R electrodes, respectively. Obviously, [$\mu_L, \mu_R$] defines a bias window (BW) for electron transmission, and we will focus on the effective states in the BW in the following discussions. More details of this approach can be obtained in the previous publications [39-41].

Figure 5(c) shows the transmission spectra of z-MoB$_2$ under biases from 0 to 1.5 V. At a low bias such as 0 V, it shows an obvious quantized characteristic, similar to graphene and other graphene-like monolayers [18, 55-57]. The equilibrium conductance of z-MoB$_2$ is 7 G$_0$ (G$_0$=$e^2/h$), which is larger than that of hydrogenated borophene (2 G$_0$) [58] due to more bands crossing the E$_F$. With the increasing of bias voltage, the transmission coefficients decrease gradually in the most energy region. It is mainly attributed to the band shift of the L and R electrodes.

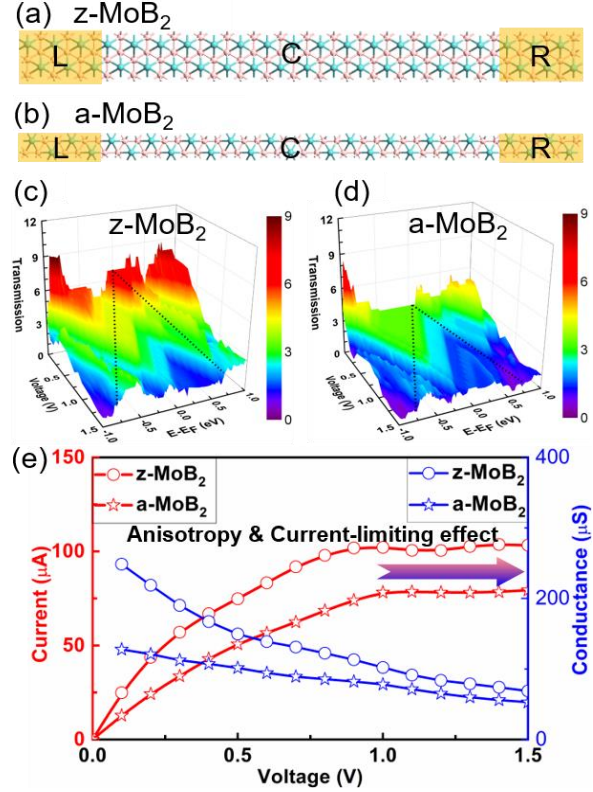

**Figure 5.** Two-probe structures of (a) z-MoB$_2$ and (b) a-MoB$_2$. $T(E,V_b)$ of (c) z-MoB$_2$ and (d) a-MoB$_2$. The dotted lines refer to the BW. The E$_F$ is shifted to zero. (e) $I-V$ and $G-V$ curves of z-MoB$_2$ and a-MoB$_2$.

Generally, the electronic transport of monolayer materials is mostly dominated by their band alignments from the inter- and intra-band transitions near the E$_F$ [59]. On the one hand, only the electrons in the states between the BW can give effective contributions to the current. On the other hand, the band characters, including the parity limitation [55], the projection of bands [18], and its localization [58], have important influence on the electron transport. As a forward bias is applied to the L and R electrodes, their bands move down and up accordingly. Some localized states such as the band A near the $\Gamma$ point (see figure 4(e)) enter the BW as the bias is beyond 1.0 V, leading to the transmission coefficients decreased significantly. Consequently, the $I-V$ curve of z-MoB$_2$ appears a current-limiting effect (see figure 5(e)), due to the decreasing transmission coefficients as the bias window expands. The saturation current is 100 µA, larger than that of hydrogenated borophene (24 µA) [58].

For the a-MoB$_2$, it has smaller transmission coefficients (see figure 5(d)) and shows obvious anisotropy as compared with z-MoB$_2$, because fewer bands cross the Fermi level along the $\Gamma$–$Y$ path (see figure 4(e)). It also shows a current-limiting effect like z-MoB$_2$ with the same mechanism but has a smaller saturation value (78 µA). The electrical anisotropy ratio $\eta = I_z/I_a$ ($I_z$ and $I_a$ are the saturation current of z-MoB$_2$ and a-MoB$_2$, respectively) is 1.3, smaller than that of χ-borophene and hydrogenated borophene (1.5) [58, 60]. Moreover, the conductances ($G$) of both z-MoB$_2$ and a-MoB$_2$ decrease as the bias increases (see the $G-V$ curves in figure 5(e)), and the electrical anisotropy ratio has a same attenuation trend.

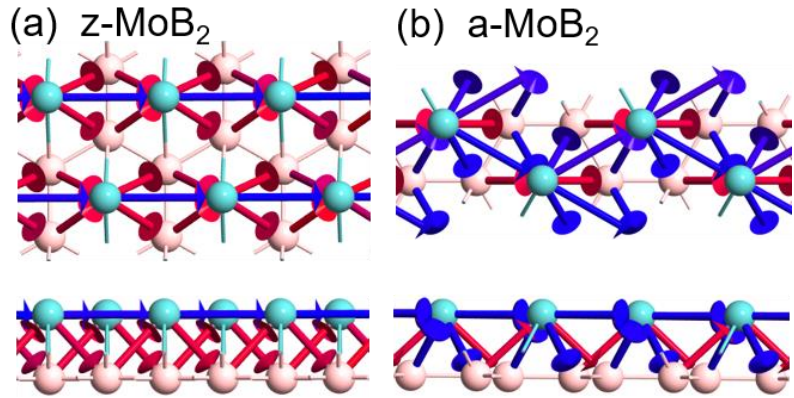

**Figure 6.** Top (upside) and side (downside) views of transmission pathways of (a) z-MoB$_2$ and (b) a-MoB$_2$ at the Fermi level under 0.5 V.

To better uncover the physical picture of electronic transport of z-MoB$_2$ and a-MoB$_2$, we plot their electron transmission pathways [61] at the $E_F$ under 0.5 V (see figure 6). For z-MoB$_2$, it has two types of local current pathways, i.e., through Mo-B (red arrows) and Mo-Mo (blue arrows) bonds (see figure 6(a)). Note that B-B bonds give little contribution to the local current pathways. This is different from hydrogenated borophene [58], whose B-B bonds dominate its local current pathways. For the case of a-MoB$_2$, it has the same local current pathways as z-MoB$_2$ but with different directions, due to their different atom lattice positions (see figure 6(b)).

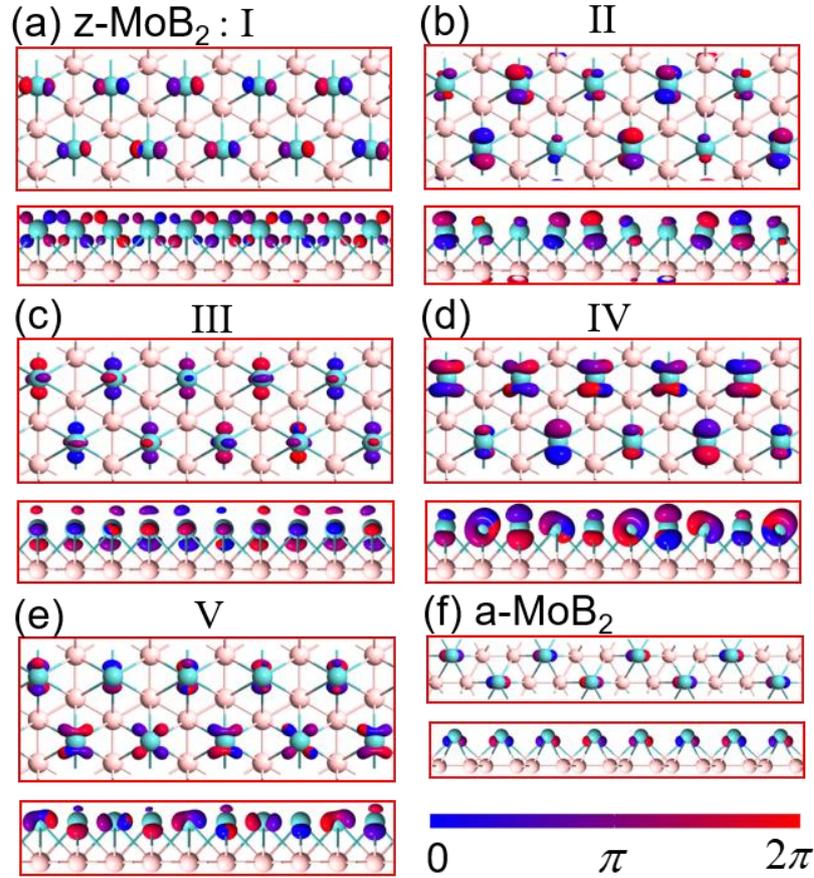

**Figure 7.** Top (upside) and side (downside) views of transmission eigenstates at the $E_F$ under 0.5 V. (a)-(e) for z-MoB$_2$, and (f) for a-MoB$_2$.

Figure 7 shows the corresponding transmission eigenstates of z-MoB$_2$ and a-MoB$_2$. For z-MoB$_2$, it has five degenerate transmission eigenstates (see figures 7(a) to (e)) contributing to the current, which are composed of hybrid Mo $d$-orbitals and primarily stem from the five bands crossing the Fermi level along the $\Gamma$–$Y$ path (see figure 4(e)). Note that the degenerate transmission eigenstates will decrease under a high bias, such as to three under 1.5 V. This is ascribed to that some bands (at $\Gamma$ point) enter the bias window at a high bias and verify its current-limiting effect. While, for a-MoB$_2$, it only has one transmission eigenstates (see figure 7(f)), suggesting its low-level conductance and saturation current.

## 4. Conclusions

In summary, we predict a new hexagonal transition metal diboride MoB$_2$ monolayer by means of first-principles calculations and confirm its the geometric structure, dynamic and thermal stabilities. Our results demonstrate that the MoB$_2$ ML has a high isotropic Young's stiffness value (641 GPa) along the zigzag and armchair directions and has an ultralow thermal conductivity (11.54 W m$^{-1}$ K$^{-1}$) at room temperature. It shows a metallic feature as many Mo states are around the Fermi level. Along the zigzag and armchair directions, its $I-V$ curve presents an obvious electrical anisotropy and a useful current-limiting effect. The present work demonstrate that the MoB$_2$ ML has several potential applications in nanoelectronics and is a good candidate of ultrathin high-strength mechanical materials, heat insulating materials, electrical-anisotropy-based materials, and current limiters. Our findings are useful for the preparation and utilization of the ultrathin 2D materials based on the transition metal diborides.

## ORCID iDs

Yipeng An https://orcid.org/0000-0001-5477-4659
Ruqian Wu https://orcid.org/ 0000-0002-6156-7874
Tianxing Wang https://orcid.org/0000-0003-3659-8801

## Acknowledgments

The work at Henan Normal University was supported by the National Natural Science Foundation of China (Grant Nos. 11774079, 61774059 and U1704136), the Scientific and Technological Innovation Program of Henan Province's Universities (Grant No. 20HASTIT026), the Young Backbone Teacher Training Program of Henan Province's Higher Education (Grant No. 2017GGJS043), the Science Foundation for the Excellent Youth

Scholars of Henan Normal University (Grant No. 2016YQ05), the Henan Overseas Expertise Introduction Center for Discipline Innovation (Grant No. CXJD2019005), and the High-Performance Computing Centre of Henan Normal University. The work at the University of California at Irvine was supported by the US DOE-BES under Grant No. DE-FG02-05ER46237.